\documentclass{PoS}
\usepackage{wrapfig}
\def\GeV{{\rm GeV}}

\title{The Effect of Final HERA inclusive Cross Section Data on MMHT2014 PDFs }

\ShortTitle{MMHT 2014 PDFs - HERA}

\author{\speaker{R.S.~Thorne} \\
        Department of Physics and Astronomy, \\  
        University College London, WC1E 6BT, UK\\
        E-mail: \email{robert.thorne@ucl.ac.uk}}

\author{L.A.~Harland-Lang\\
        Department of Physics and Astronomy, \\  
        University College London, WC1E 6BT, UK\\
        E-mail: \email{l.harland-lang@ucl.ac.uk}}

\author{A.D.~Martin\\
        Institute for Particle Physics Phenomenology,\\
        University of Durham, DH1 3LE, UK \\ 
        E-mail: \email{A.D.Martin@durham.ac.uk}}

\author{P.~Motylinski \\
        Department of Physics and Astronomy, \\  
        University College London, WC1E 6BT, UK\\
        E-mail: \email{p.motylinski@ucl.ac.uk}}

\abstract{We investigate the effect of including the HERA run I + II 
combined cross section data on the MMHT2014 PDFs. We present the fit quality
within the context of the global fit and when only the HERA data are included.  
We examine the changes in both the central values and uncertainties in the 
PDFs. We find that the prediction for the data is good, and only relatively 
small improvements in $\chi^2$ and changes in the PDFs are obtained 
with a refit at both 
NLO and NNLO. PDF uncertainties are slightly reduced. There is a small
dependence of the fit quality on the value of $Q^2_{\min}$.  }

\FullConference{The European Physical Society Conference on High Energy Physics\\
		22--29 July 2015\\
		Vienna, Austria}

\begin{document}

\vspace{-0.0cm}

The MSTW2008 PDFs \cite{Martin:2009iq} were recently updated with a fit in 
the same general framework -- the MMHT2014 PDFs \cite{Harland-Lang:2014zoa}. 
These were an improvement due to a number of developments in theory
or procedures. For example, we now use extended parameterisation 
with Chebyshev
polynomials, and freedom in deuteron nuclear corrections 
introduced in \cite{Martin:2012da}, which led to a change in
the $u_V-d_V$ distribution.
We use the ``optimal'' GM-VFNS choice \cite{Thorne:2012az}
which is smoother near
to heavy flavour transition points, particularly at NLO.
Correlated systematic uncertainties are now treated as multiplicative not 
additive. We have also changed the 
value of the charm branching ratio to muons used to $B_{\mu} = 0.092 \pm 10\%$ 
\cite{Bolton:1997pq}.
There are also a wide variety of
new data sets included in the fit. 
This includes 
LHC (ATLAS, CMS and LHCb) $W,Z$ cross sections
differential in rapidity, and Drell Yan at high and low mass, and 
also data on $\sigma_{t\bar t}$ from the Tevatron and from ATLAS and CMS.
At NLO we also include ATLAS and CMS inclusive jet data, though 
do not yet include these data at NNLO. Previous
analyses have used threshold corrections for Tevatron data and we continue 
to include these data,  but at 
the LHC we are often far from threshold.   
The full 
NNLO calculation \cite{Ridder:2013mf,Currie:2013dwa} is nearing 
completion.

\begin{figure}[]
\vspace{-0.2cm}
\centerline{\includegraphics[width=0.6\textwidth]{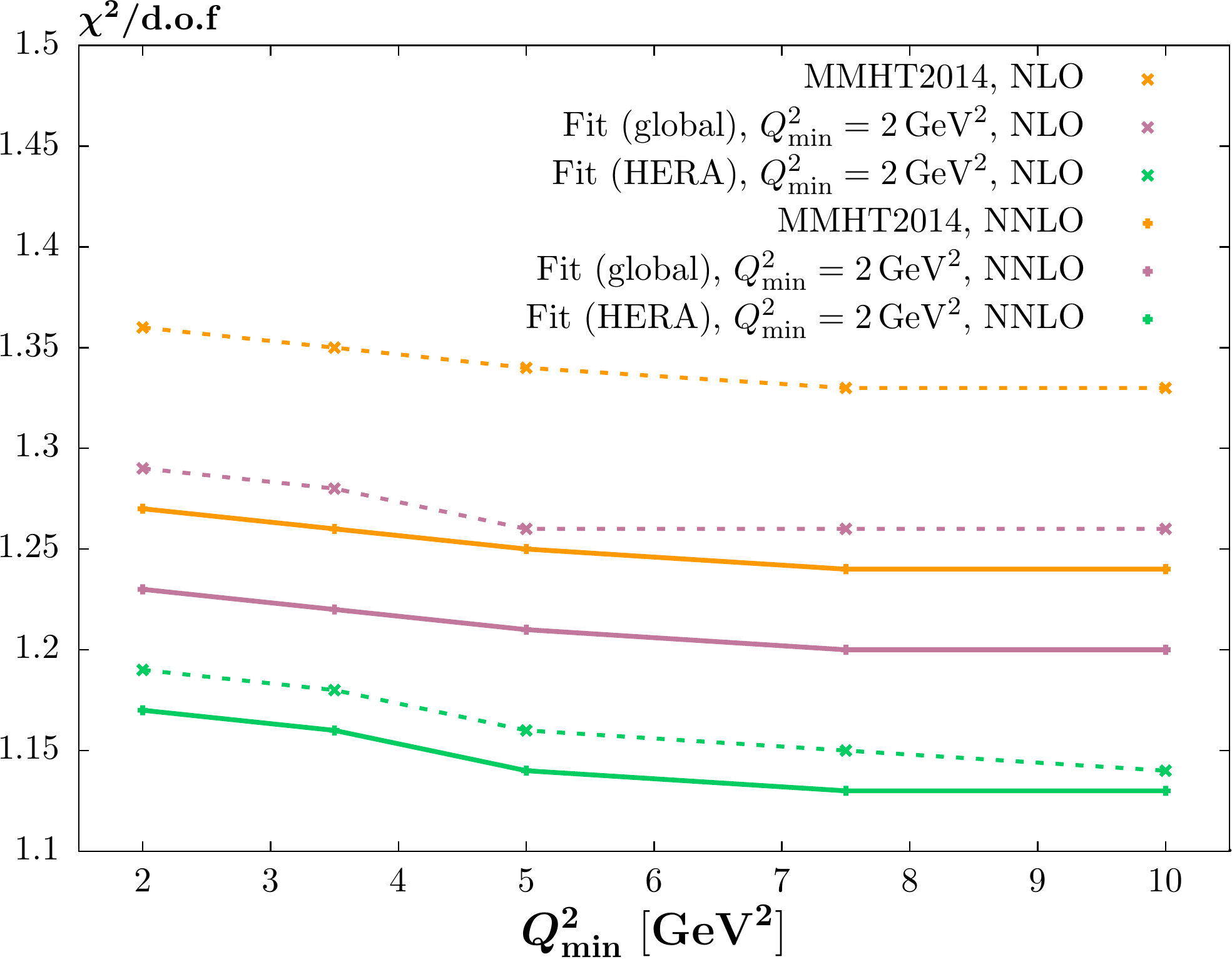}}
\vspace{-0.1cm}
\caption{A comparison of $\chi^2$ per point for 
the three variations of NLO and NNLO fits, all with
fixed lower $Q^2_{\min}=2~\GeV^2$, for different $Q^2_{\min}$ on the HERA
data used in the calculations.}
\vspace{-0.2cm}
\label{Fig1} 
\end{figure}

There are also various changes in non-LHC data sets, e.g. we include some updated Tevatron asymmetry data sets. 
The single most important change in data included is the 
replacement of the HERA run I neutral and charged current data provided 
separately by H1 and ZEUS with the combined HERA data set \cite{Aaron:2009aa}
(we also include HERA combined data on $F_2^c(x,Q^2)$
\cite{Abramowicz:1900rp}), which is the data set which provides
the best single constraint on PDFs, particularly the gluon at all 
$x < 0.1$. 
However, we decided not to include any separate run II H1 and ZEUS data sets, 
since it was clear the run I $+$ II combination data would soon 
appear. This has now happened, and the data, and the accompanying 
PDF analysis, are in \cite{Abramowicz:2015mha}. 
It was left as an open question in \cite{Harland-Lang:2014zoa} when 
an update of MMHT2014 PDFs would be required. Significant new LHC data would 
be one potential reason, and the full NNLO calculation of the jet
cross sections might be another. However, the potential impact of the final 
HERA inclusive cross section data was another factor, 
it  being possible that these 
alone might produce a significant change in the central value of the PDFs
and their uncertainties. Hence, it is now obviously a high priority to 
investigate their impact. 

\begin{figure}[]
\vspace{-0.2cm}
\centerline{\includegraphics[width=0.48\textwidth]{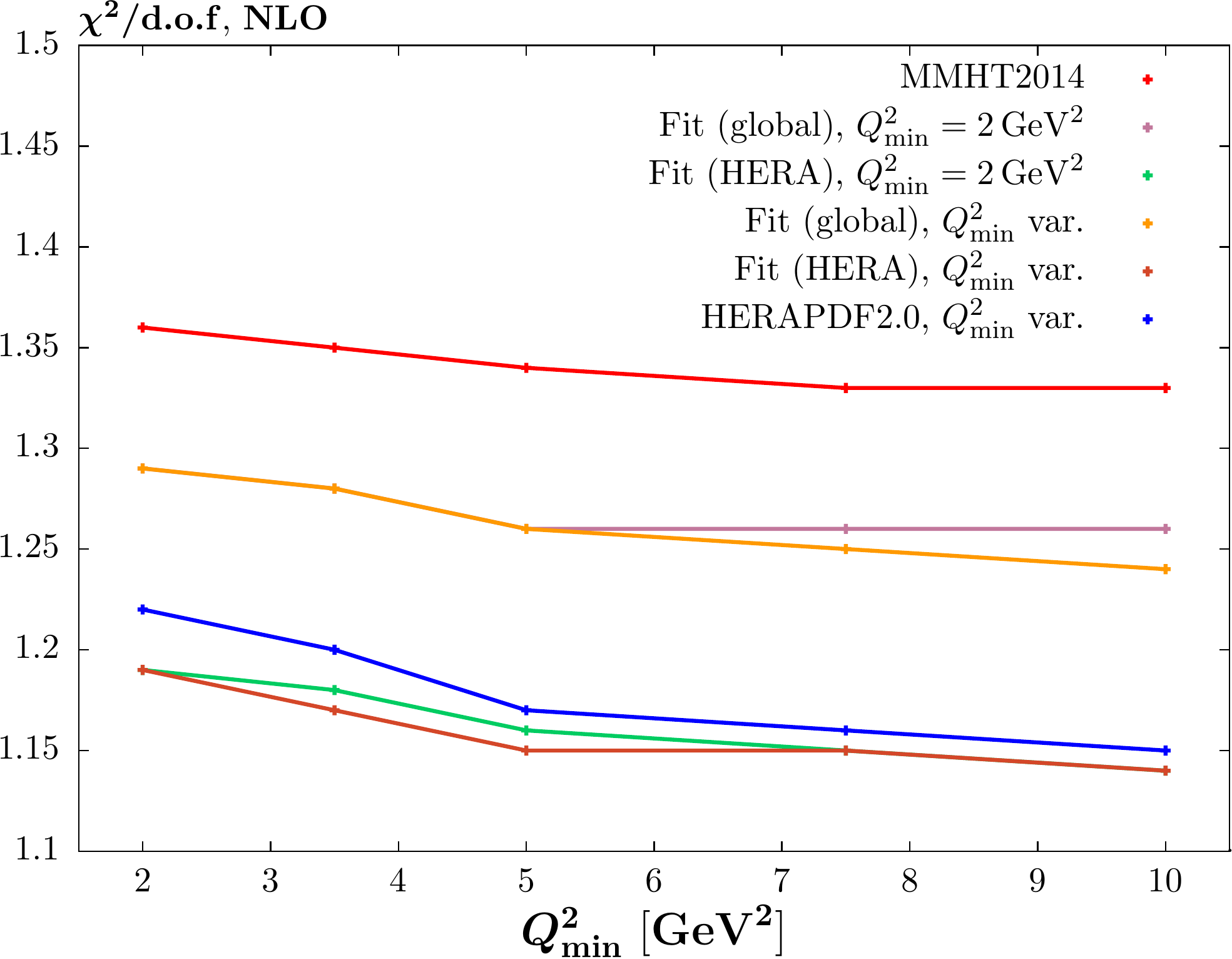}
\includegraphics[width=0.48\textwidth]{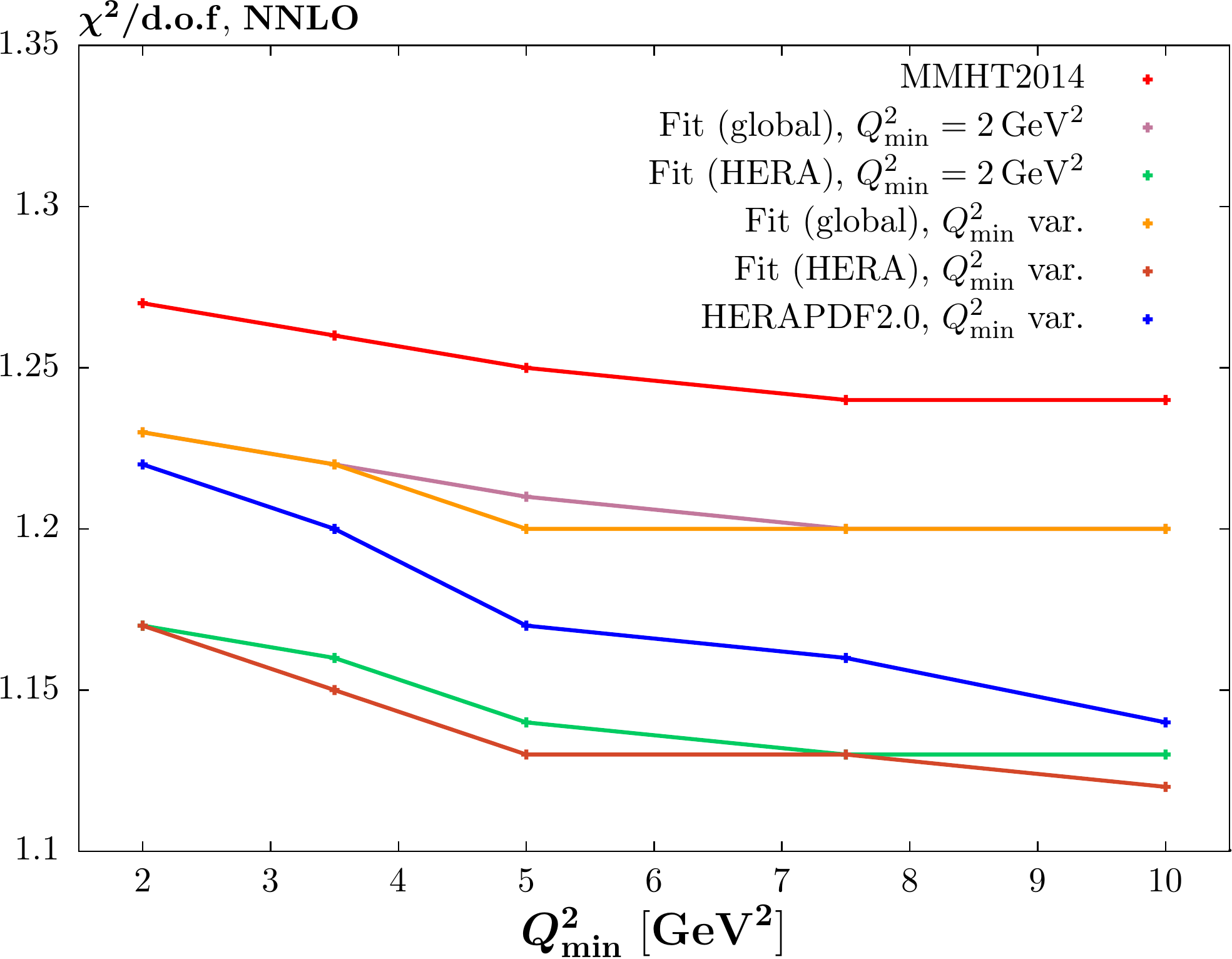}}
\vspace{-0.2cm}
\caption{The variation in $\chi^2$ per point when increasing the 
$Q^2_{\min}$ cut, 
on only HERA Run I $+$ II data, at both NLO (left) and  at NNLO (right). }
\vspace{-0.2cm}
\label{Fig2} 
\end{figure}

Using our standard cut of $Q^2_{\min}=2~\GeV^2$ there are 1185 HERA 
data points 
with 162 correlated systematics and 7 procedural uncertainties. 
These are separated into 7 subsets, depending on whether it is $e^+$
or $e^-$ scattering from the proton, whether it is neutral or charged 
current scattering and on the proton beam energy $E_p$. 
This is to be compared to 621 data points, separated into 5 subsets, 
with generally larger uncertainties, from the HERA I combined data used 
previously (though this has fewer correlated systematics).  
We first simply investigate the fit quality from the predictions 
using MMHT2014 PDFs.\footnote{Our numbers are very 
slightly different to those presented at EPS2015 due to 
the correction of a minor bug.}  This is already rather good:

\medskip

\noindent\centerline{$\chi^2_{NLO} = 1611/1185 = 1.36$ per point.}

\noindent\centerline{$\chi^2_{NNLO} = 1503/1185 = 1.27$ per point.}

\medskip

\noindent In contrast HERAPDF2.0 PDFs, which are fit to (only) these data  
obtain $\sim 1.20$ per point with $Q^2_{\min}=2~\GeV^2$ at both NLO and NNLO.
Next we try refitting in the context of our standard global fit, i.e. we 
simply replace the previous HERA run I data with the 
new run I $+$ II combined data. The fit improves to   

\medskip

\noindent\centerline{$\chi^2_{NLO} = 1533/1185 = 1.29$ per point, with
deterioration $\Delta \chi^2 = 29$ in other data.} 

\noindent\centerline{$\chi^2_{NNLO} = 1457/1185 = 1.23$ per point, with
deterioration $\Delta \chi^2 = 12$ in other data.} 

\medskip

\noindent This is a significant, but hardly dramatic improvement (less than
the improvement after refitting when HERA run I combined data were first 
introduced into the MSTW2008 fitting framework), i.e. the MMHT2014 PDFs were
already giving quite close to the best fit within the global fit framework.  
In order to compare more directly with the HERAPDF2.0 study we
also fit only HERA run I $+$ II data. This requires us to 
fix 4 of our PDF parameters in order to avoid particularly unusual PDFs, in
practice a very peculiar, and potentially pathological strange distribution --
HERA data not having any direct constraint on this. We allow the strange
+ antistrange distribution to have a free normalisation and high-$x$ power, 
but all other shape freedom is removed. The strange-antistrange asymmetry is 
fixed to the MMHT2014 default values. The result is  

\medskip

\noindent\centerline{$\chi^2_{NLO} = 1416/1185 = 1.19$ per point}

\noindent\centerline{$\chi^2_{NNLO} = 1381/1185 = 1.17$ per point}
  
\medskip

\noindent Hence, in this case, as well as the global fit, the NNLO fit 
quality is definitely better than NLO.
Part of this is due to the charged current data $\chi^2$ which is $~$ 20 
units better in HERA data only fits than the global fits and $~$ 10 units 
better at NNLO than at NLO. There appears to be some tension between 
these charged
current data and other data in the global fit, though this is partially 
resolved at NNLO. 


\begin{table}[h]
\begin{center}
\begin{tabular}{|l|c|c|c|c|c|}
\hline
          & no. points    & NLO $\chi^2_{HERA}$  &   NLO $\chi^2_{global}$ &
  NNLO $\chi^2_{HERA}$  &   NNLO $\chi^2_{global}$         \\
\hline
correlated penalty      &       & 79.9 &  113.6& 73.0 &  92.1  \\
CC $e^+p$                &  39   & 43.4 & 47.6  & 42.2 & 48.4     \\
CC $e^-p$                &  42   & 52.6 & 70.3  & 47.0 & 59.3     \\
NC $e^-p$ $E_p=920~\GeV$ &  159  & 213.6 & 233.1 & 213.5 & 226.7     \\
NC $e^+p$ $E_p=920~\GeV$ &  377  & 435.2 & 470.0 & 422.8 & 450.1   \\
NC $e^+p$ $E_p=820~\GeV$ &   70  & 67.6  & 69.8 & 71.2  & 69.5   \\
NC $e^-p$ $E_p=460~\GeV$ &   254 & 228.7 & 233.6 & 229.1 & 231.8     \\
NC $e^-p$ $E_p=920~\GeV$ &   204 & 221.6 & 228.1 & 220.2 & 225.6     \\
\hline                                                        
total      &      1145          &1342.6 &1466.1&1319.0 &1403.5  \\
\hline
    \end{tabular}
\vspace{-0.0cm}
\caption{\label{tab:t1} The $\chi^2$ for each subset of HERA I + II 
data for the four variations of fit for $Q^2_{\min}=3.5~\GeV^2$.}
\vspace{-0.4cm}
\end{center}
\end{table}

The HERAPDF2.0 analysis sees a distinct improvement in $\chi^2$ per point with 
a raising of the $Q^2_{\min}$ value for the data fit. Hence, we also 
investigate the variation of the fit quality with $Q^2_{\min}$. However, first 
we simply calculate the quality of the comparison to data as a function of 
$Q^2_{\min}$ at NLO and at NNLO without refitting, i.e. the PDFs are obtained
with the default $Q^2_{\min}=2~\GeV^2$ cut. This is shown in Fig. \ref{Fig1}
where we show a comparison of $\chi^2$ per point for 
the three variations of NLO and NNLO comparisons, i.e. MMHT2014 prediction, 
global refit with new HERA data and refit with only HERA run I $+$ II
combined data. Here it is clear that NNLO is superior, but this is less 
distinct in the refits, particularly the fit to 
only HERA data. There is a reasonable lowering of the $\chi^2$ per point 
as $Q^2_{\min}$ increases.

\begin{figure}[]
\vspace{-0.2cm}
\centerline{\includegraphics[width=0.48\textwidth]{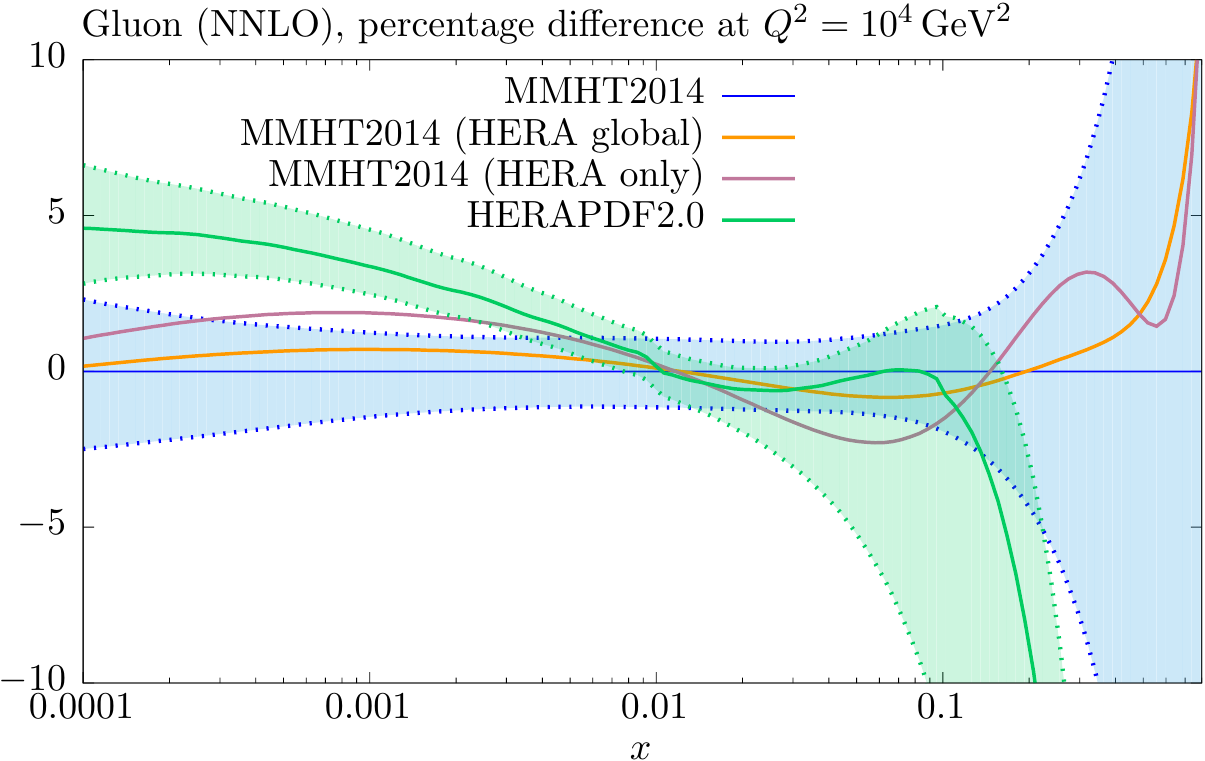}
\includegraphics[width=0.48\textwidth]{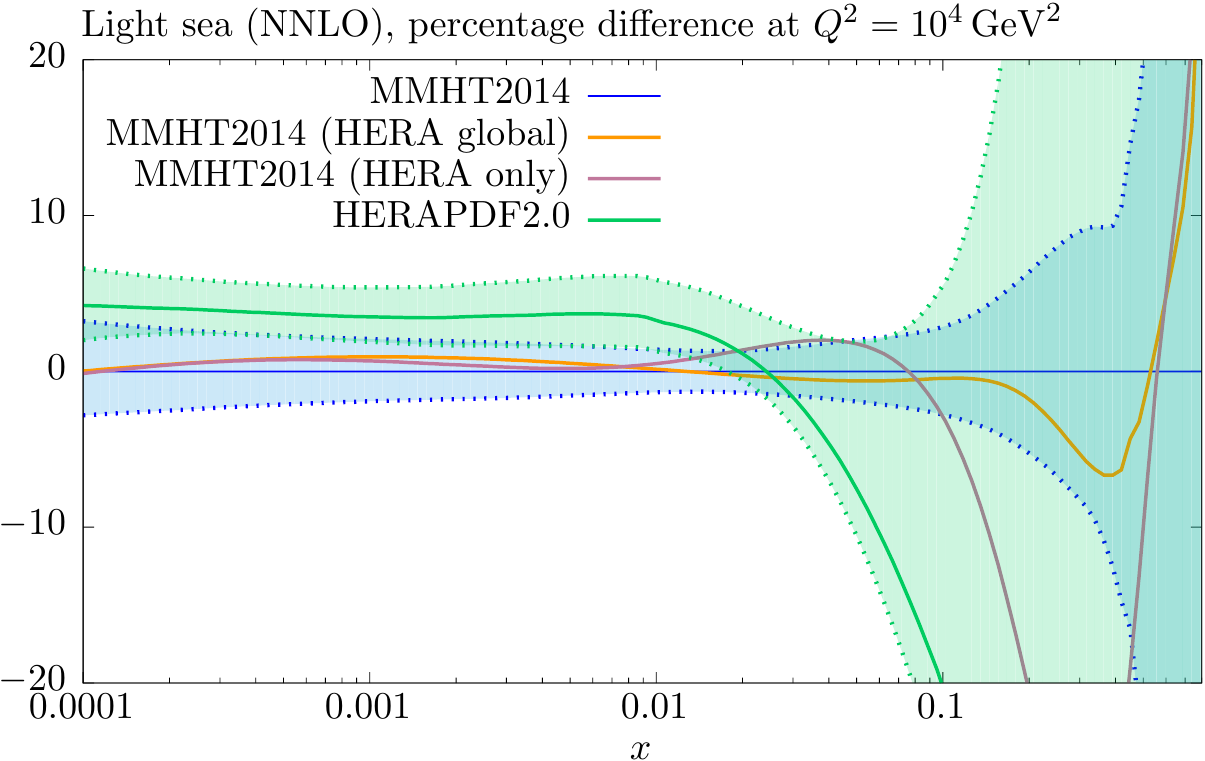}}
\centerline{\includegraphics[width=0.48\textwidth]{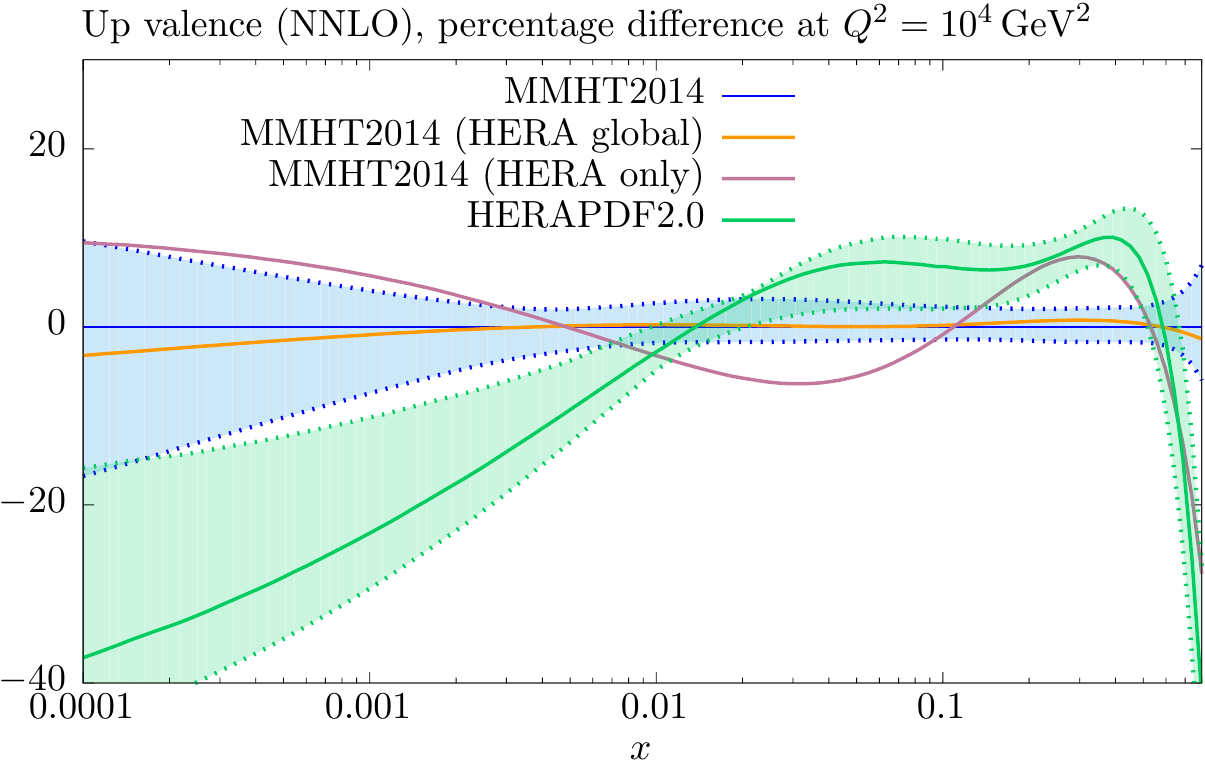}
\includegraphics[width=0.48\textwidth]{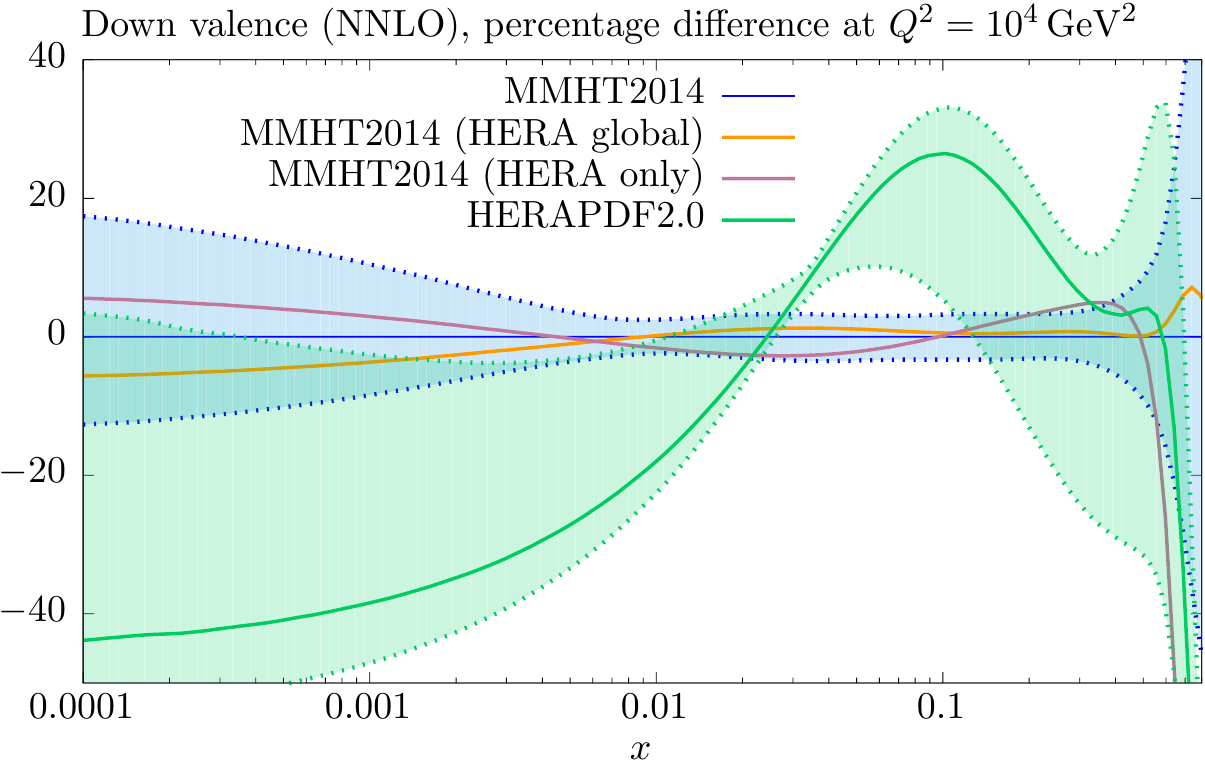}}
\vspace{-0.1cm}
\caption{The comparison of NNLO MMHT PDFs containing the new HERA data to
MMHT2014 PDFs and the HERAPDF2.0 PDFs.}
\vspace{-0.2cm}
\label{Fig3} 
\end{figure}

We also look at the effect of changing the $Q^2$ cut in the fit (on only 
the HERA combined data), 
at both NLO and NNLO. This is shown in Fig.~\ref{Fig2}, where we also show 
the trend for the HERAPDF2.0 analysis \cite{Abramowicz:2015mha}.\footnote{The 
definition of $\chi^2$ for HERAPDF2.0 is not identical, though this should be 
a very small effect.} For comparison we also include the curves for the 
$\chi^2$ per point with varying $Q^2_{\min}$ but with the fits performed for 
$Q^2_{\min}=2~\GeV^2$.  In fact we note that while there is an improvement 
in $\chi^2$ per point with increasing $Q^2_{\min}$, this is very largely
achieved without any refitting. There is also less improvement in our 
analysis than for HERAPDF2.0, particularly in the 
global fit and at NNLO. In order to investigate the source of the improvement 
with increasing $Q^2_{\min}$ we looked in more detail at the fit quality for 
low $Q^2$ bins.  We found that there was quite a large degree of 
point-to-point fluctuation 
in theory/data at low $Q^2$ rather than any 
obvious systematic issue (though the 
lowest $x$ point at each $Q^2$ is often below theory, i.e. there is 
limited evidence of more of a turn-over in data than theory). 
In many cases it seems impossible to 
avoid some rather high $\chi^2$ points with any smooth curve.  
We present the detail of the breakdown of $\chi^2$ into each of the data 
subsets in Table~\ref{tab:t1}. We do this for $Q^2_{\min}=3.5~\GeV^2$ to allow 
direct comparison to the HERAPDF2.0 results, and also for possible comparison 
with e.g. NNPDF and CT results.

In Fig.~\ref{Fig3} we show the central values of the NNLO PDFs from the fits
including the new HERA combined data, comparing them to MMHT2014 PDFs (with 
uncertainties) and the HERAPDF2.0 PDFs (also with uncertainties). The modified 
PDFs are always very well within the MMHT2014  
uncertainty bands. Indeed, the predictions for e.g. $gg \to H$ change by 
$< 0.2\%$ for the full range of LHC energies.
The PDFs from the HERA run I $+$ II data only fit are 
in some ways similar to those of HERAPDF2.0, e.g. the up valence 
quark for $x>0.2$, which show some significant 
deviations to the two global fits PDF sets. However, the features are not 
universal, with the gluon and the down valence being much more similar to 
MMHT2014 than HERAPDF2.0. This may be a feature of the differing 
parameterisations used in the two studies.   

\begin{figure}[]
\vspace{-0.2cm}
\centerline{\includegraphics[width=0.48\textwidth]{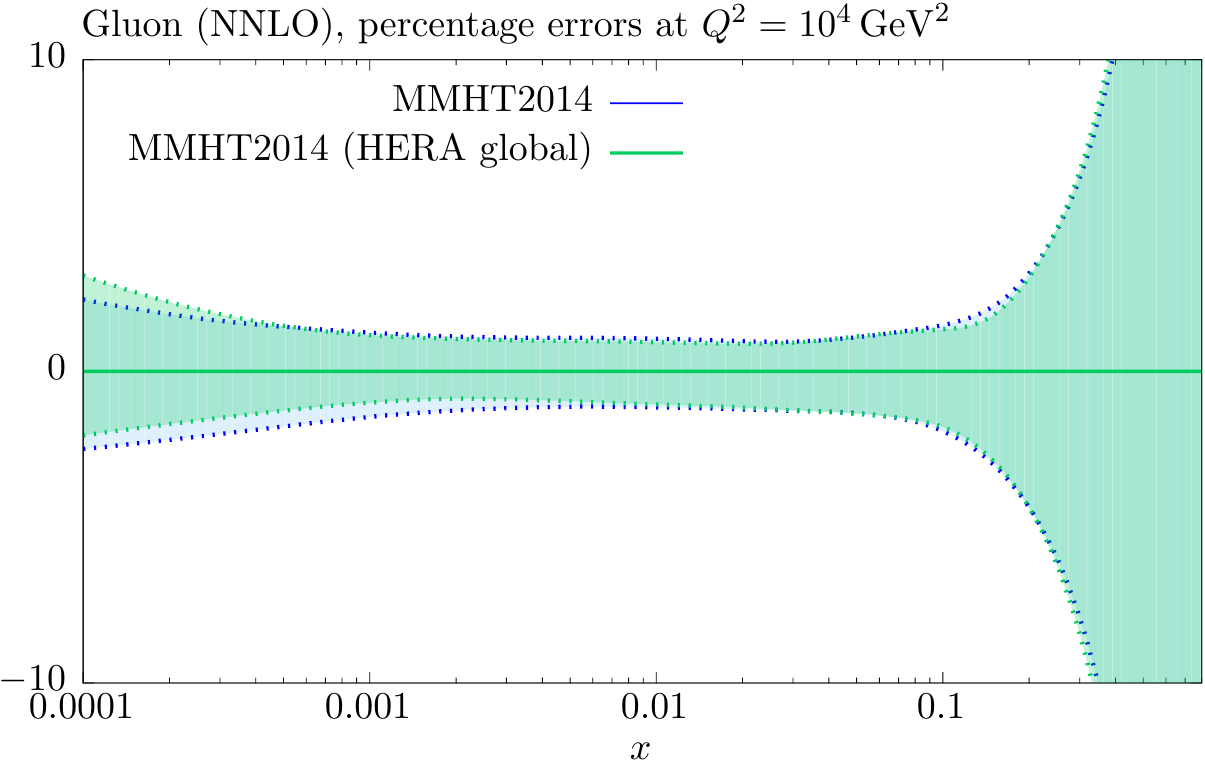}
\includegraphics[width=0.48\textwidth]{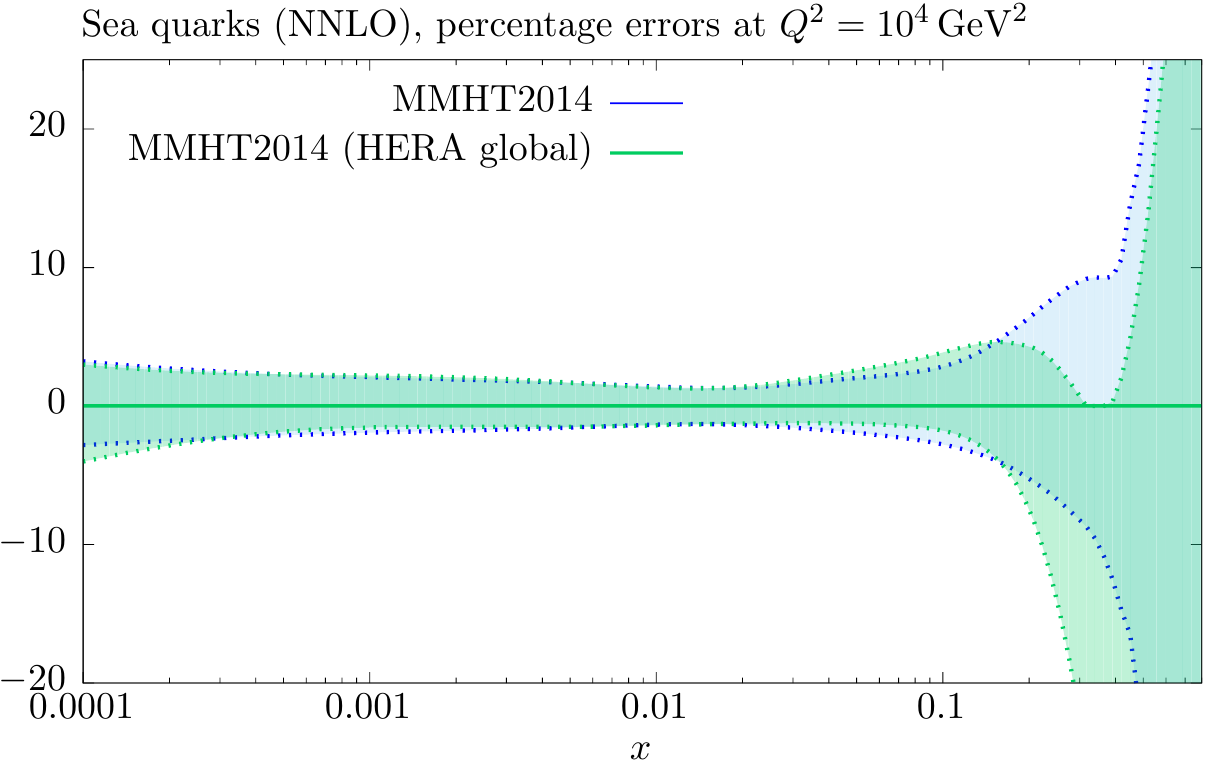}}
\centerline{\includegraphics[width=0.48\textwidth]{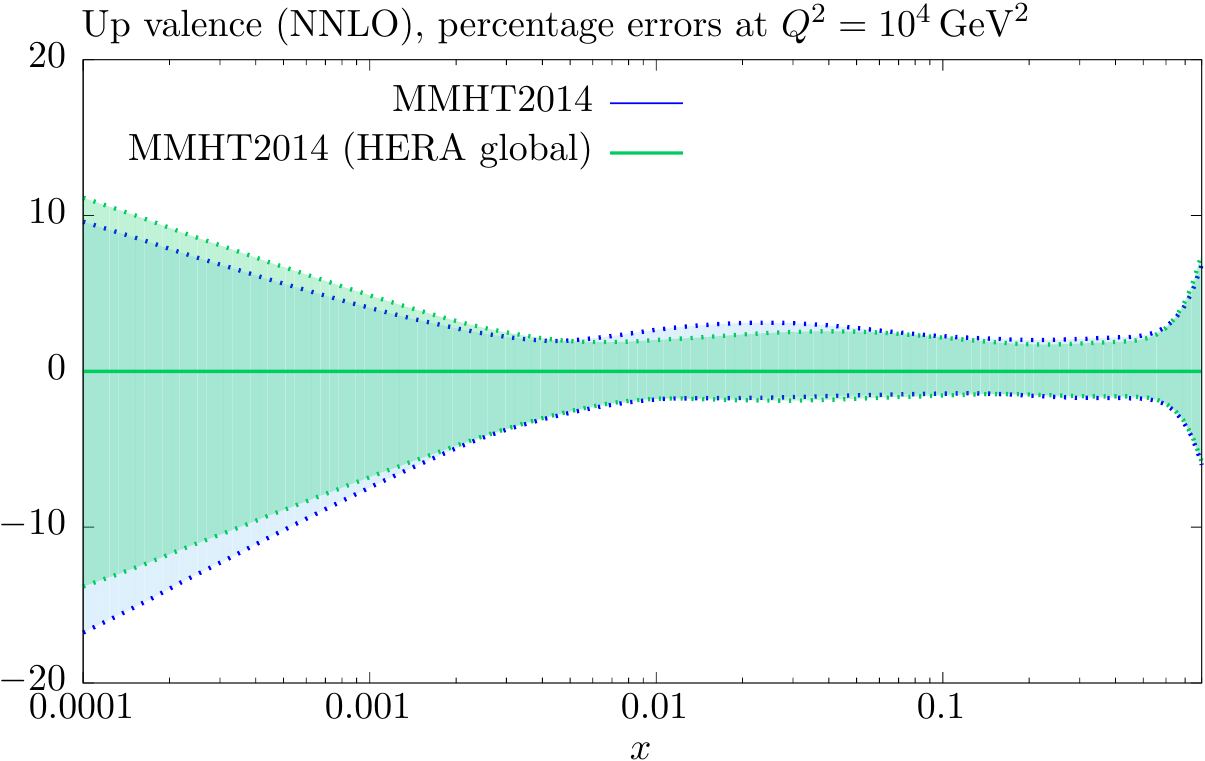}
\includegraphics[width=0.48\textwidth]{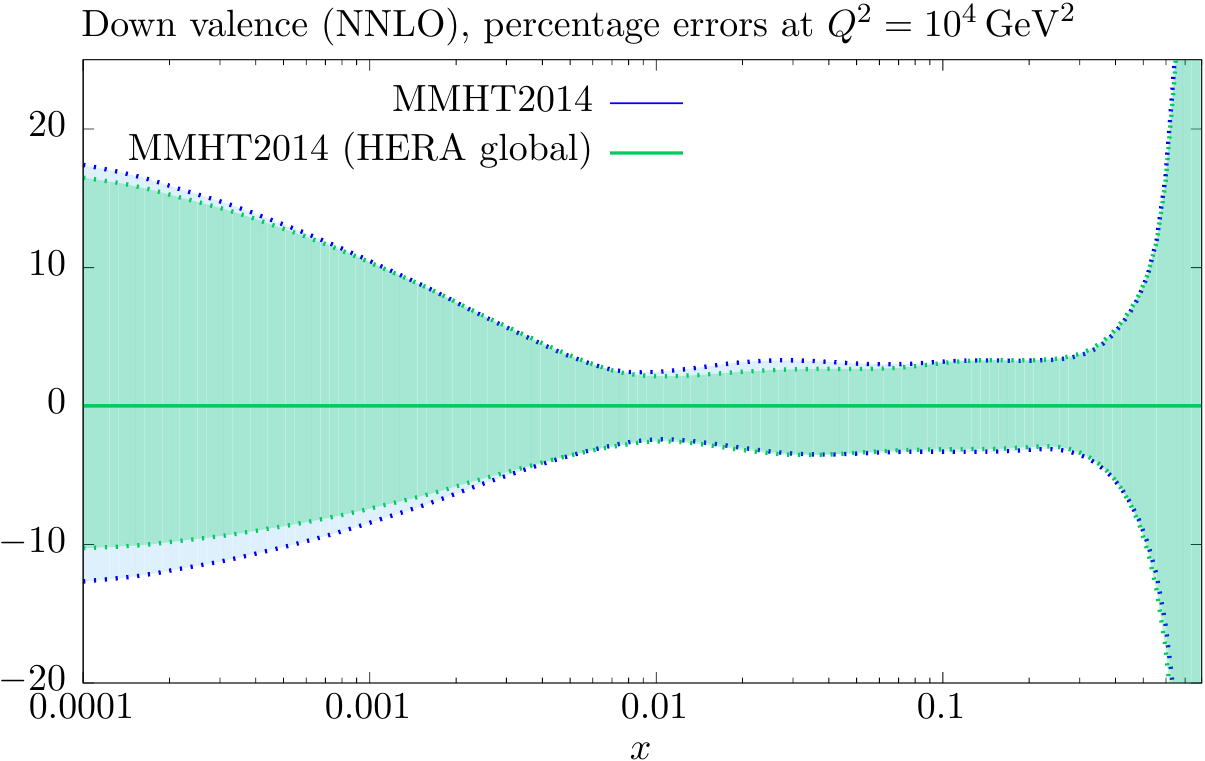}}
\vspace{-0.1cm}
\caption{The comparison of the uncertainty on NNLO MMHT PDFs from a global 
fit containing the 
new HERA data to that on the MMHT2014 PDFs. }
\vspace{-0.2cm}
\label{Fig4} 
\end{figure}

We also investigate the effect of the new data on the uncertainties of
the PDFs. 
In Fig.~\ref{Fig4} we compare the uncertainties for the NNLO PDFs
including the HERA run I $+$ II data to the MMHT2014 PDFS. These 
are quite similar to MMHT2014 in most features.
The most obvious improvement is to the gluon for $x < 0.01$. 
There is also slight improvement in some places for the valence quarks. 
We understand that NNPDF results on central values and uncertainties 
are similar \cite{NNPDF}. Hence, 
the HERA run I $+$ II combined data gives us our single best constraint 
on PDFs, and determines the gluon at low $x$ even more accurately than 
before, but its inclusion does not suggest updates of existing PDFs are  
immediately necessary. 

\vspace{-0.3cm}

\section*{Acknowledgements}

\vspace{-0.3cm}

We would like to thank A.M Cooper-Sarkar, A. Geiser and V. Radescu
for discussions on the HERA data and PDFs, and S. Forte and J. Rojo on 
NNPDF results. This work is
supported partly by the London Centre for Terauniverse Studies (LCTS),
using funding from the European Research Council via the Advanced 
Investigator Grant 267352. We thank the
Science and Technology Facilities Council (STFC) for support via grant
awards ST/J000515/1 and ST/L000377/1.

\vspace{-0.4cm}


\end{document}